\documentclass[12pt]{article}
\usepackage{amsfonts}
\usepackage{amssymb}

\def\hybrid{\topmargin -20pt    \oddsidemargin 0pt
        \headheight 0pt \headsep 0pt
        \textwidth 6.35in       
        \textheight 9.25in       
        \marginparwidth .875in
        \parskip 5pt plus 1pt   \jot = 1.5ex}

\hybrid

\def\baselinestretch{1.2}

\catcode`\@=11

\def\marginnote#1{}
\newcount\hour
\newcount\minute
\newtoks\amorpm
\hour=\time\divide\hour by60
\minute=\time{\multiply\hour by60 \global\advance\minute by-\hour}
\edef\standardtime{{\ifnum\hour<12 \global\amorpm={am}%
        \else\global\amorpm={pm}\advance\hour by-12 \fi
        \ifnum\hour=0 \hour=12 \fi
        \number\hour:\ifnum\minute<10 0\fi\number\minute\the\amorpm}}
\edef\militarytime{\number\hour:\ifnum\minute<10 0\fi\number\minute}

\def\draftlabel#1{{\@bsphack\if@filesw {\let\thepage\relax
   \xdef\@gtempa{\write\@auxout{\string
      \newlabel{#1}{{\@currentlabel}{\thepage}}}}}\@gtempa
   \if@nobreak \ifvmode\nobreak\fi\fi\fi\@esphack}
        \gdef\@eqnlabel{#1}}
\def\@eqnlabel{}
\def\@vacuum{}
\def\draftmarginnote#1{\marginpar{\raggedright\scriptsize\tt#1}}

\def\draft{\oddsidemargin -.5truein
        \def\@oddfoot{\sl preliminary draft \hfil
        \rm\thepage\hfil\sl\today\quad\militarytime}
        \let\@evenfoot\@oddfoot \overfullrule 3pt
        \let\label=\draftlabel
        \let\marginnote=\draftmarginnote
   \def\@eqnnum{(\theequation)\rlap{\kern\marginparsep\tt\@eqnlabel}%
\global\let\@eqnlabel\@vacuum}  }

\def\preprint{\twocolumn\sloppy\flushbottom\parindent 2em
        \leftmargini 2em\leftmarginv .5em\leftmarginvi .5em
        \oddsidemargin -.5in    \evensidemargin -.5in
        \columnsep .4in \footheight 0pt
        \textwidth 10.in        \topmargin  -.4in
        \headheight 12pt \topskip .4in
        \textheight 6.9in \footskip 0pt
        \def\@oddhead{\thepage\hfil\addtocounter{page}{1}\thepage}
        \let\@evenhead\@oddhead \def\@oddfoot{} \def\@evenfoot{} }

\def\numberbysection{\@addtoreset{equation}{section}
        \def\theequation{\thesection.\arabic{equation}}}

\def\underline#1{\relax\ifmmode\@@underline#1\else
        $\@@underline{\hbox{#1}}$\relax\fi}

\def\titlepage{\@restonecolfalse\if@twocolumn\@restonecoltrue\onecolumn
     \else \newpage \fi \thispagestyle{empty}\c@page\z@
        \def\thefootnote{\fnsymbol{footnote}} }

\def\endtitlepage{\if@restonecol\twocolumn \else \newpage \fi
        \def\thefootnote{\arabic{footnote}}
        \setcounter{footnote}{0}}  

\catcode`@=12
\relax

\def\figcap{\section*{Figure Captions\markboth
        {FIGURECAPTIONS}{FIGURECAPTIONS}}\list
        {Figure \arabic{enumi}:\hfill}{\settowidth\labelwidth{Figure
999:}
        \leftmargin\labelwidth
        \advance\leftmargin\labelsep\usecounter{enumi}}}
 \relax
\def\tablecap{\section*{Table Captions\markboth
        {TABLECAPTIONS}{TABLECAPTIONS}}\list
        {Table \arabic{enumi}:\hfill}{\settowidth\labelwidth{Table
999:}
        \leftmargin\labelwidth
        \advance\leftmargin\labelsep\usecounter{enumi}}}
 \relax
\def\reflist{\section*{References\markboth
        {REFLIST}{REFLIST}}\list
        {[\arabic{enumi}]\hfill}{\settowidth\labelwidth{[999]}
        \leftmargin\labelwidth
        \advance\leftmargin\labelsep\usecounter{enumi}}}
 \relax
\makeatletter
\newcounter{pubctr}
\def\publist{\@ifnextchar[{\@publist}{\@@publist}}
\def\@publist[#1]{\list
        {[\arabic{pubctr}]\hfill}{\settowidth\labelwidth{[999]}
        \leftmargin\labelwidth
        \advance\leftmargin\labelsep
        \@nmbrlisttrue\def\@listctr{pubctr}
        \setcounter{pubctr}{#1}\addtocounter{pubctr}{-1}}}
\def\@@publist{\list
        {[\arabic{pubctr}]\hfill}{\settowidth\labelwidth{[999]}
        \leftmargin\labelwidth
        \advance\leftmargin\labelsep
        \@nmbrlisttrue\def\@listctr{pubctr}}}
 \relax
\makeatother
\newskip\humongous \humongous=0pt plus 1000pt minus 1000pt

\newif\ifdtup

\relax

\def\be{\begin{equation}}
\def\ee{\end{equation}}
\def\ba{\begin{eqnarray}}
\def\ea{\end{eqnarray}}

\def\r{\rho}
\def\a{\alpha}

\def\b{\beta}

\def\g{\gamma}
\def\G{\Gamma}
\def\d{\delta}

\def\e{\epsilon}

\def\th{\theta}

\def\om{\omega}
\def\Om{\Omega}
\def\l{\lambda}

\def\s{\sigma} 
\def\S{\Sigma}

\def\bs{\bigskip}

\def\qq{\qquad}

\def\IR{\relax{\rm I\kern-.18em R}}
\def\II{\relax{\rm 1\kern-.35em1}}

\def \ha {{1\over 2}}

\def \ov {\over}

\def\IR{\relax{\rm I\kern-.18em R}}
\def\inv{^{\raise.15ex\hbox{${\scriptscriptstyle -}$}\kern-.05em 1}}

\def\hi{{\hat i}}
\def\hj{{\hat j}}
\def\hk{{\hat k}}

\begin{document}


\newcommand{\beq}{\begin{equation}}
\newcommand{\eeq}[1]{\label{#1}\end{equation}}
\newcommand{\ber}{\begin{eqnarray}}
\newcommand{\eer}[1]{\label{#1}\end{eqnarray}}
\newcommand{\eqn}[1]{(\ref{#1})}
\begin{titlepage}
\begin{center}

\hfill NEIP-02-009\\
\vskip -.1 cm
\hfill hep--th/0211130\\

\vskip .5in

{\Large \bf Holonomy from wrapped branes}

\vskip 0.4in

{\bf Rafael Hern\'andez$^1$}\phantom{x} and\phantom{x}
 {\bf Konstadinos Sfetsos}$^2$ 
\vskip 0.1in

${}^1\!$
Institut de Physique, Universit\'e de Neuch\^atel\\
Breguet 1, CH-2000 Neuch\^atel, Switzerland\\
{\footnotesize{\tt rafael.hernandez@unine.ch}}

\vskip .2in

${}^2\!$
Department of Engineering Sciences, University of Patras\\
26110 Patras, Greece\\
{\footnotesize{\tt sfetsos@mail.cern.ch, des.upatras.gr}}\\

\end{center}

\vskip .3in

\centerline{\bf Abstract}

Compactifications of M-theory on manifolds with reduced holonomy arise as 
the local eleven-dimensional description of D6-branes wrapped on 
supersymmetric cycles 
in manifolds of lower dimension with a 
different holonomy group. 
Whenever the isometry group $SU(2)$ is present, eight-dimensional gauged 
supergravity is a natural arena for such investigations. 
In this paper we use this approach and review the eleven dimensional description of 
D6-branes wrapped on coassociative 4-cycles, on deformed 3-cycles 
inside Calabi--Yau threefolds and on K\"ahler 4-cycles.

\noindent

\vskip 2.2in
\noindent

\small{Proceedings of the RTN Workshop {\em{``The quantum structure of 
spacetime and the geometric 
nature of fundamental interactions''}}, 
Leuven, September 2002. Based on a talk given by R. H.}

\end{titlepage}
\vfill
\eject

\def\baselinestretch{1.2}


\baselineskip 18pt

Gravity duals of field theories with low supersymmetry can be constructed 
wrapping branes 
on supersymmetric cycles. The resulting field theories are twisted since 
preserving some ammount of 
supersymmetry after wrapping the brane requires relating 
the spin connection on the cycle with some external R-symmetry gauge fields. 
The idea is quite simple: a supersymmetric theory on a curved manifold 
$\Sigma$ will break supersymmetry because in general it will not be 
possible to find a covariantly constant spinor satisfying
$(\partial_{\mu}+\omega_{\mu}(\Sigma)) \epsilon = 0$. 
However, in the presence of a global R-symmetry 
an external gauge field can be coupled to the R-current and constant spinors 
are covariantly constant as well. The coupling to the external 
field exchanges the 
spins, resulting into a twisted theory \cite{BSV}, i.e. 
to preserve supersymmetry branes must wrap a supersymmetric cycle. 
These twists can be naturally performed within gauged supergravities \cite{MN}
(see \cite{CV} for earlier related work), 
and may involve in a quite non-trivial way the scalar fields in the theory 
\cite{HS}. 
The dual supergravity solutions describing branes of diverse dimensions 
wrapped on various supersymmetric cycles 
can be naturally constructed in an appropriate gauged supergravity, 
and eventually 
lifted to ten or eleven dimensions, an approach that has been applied to 
wrapped D6-branes \cite{EN}-\cite{D6},\cite{HS}, and also has been used to 
further develop the study of wrapped fivebranes 
\cite{D5} and to 
obtain solutions for branes of dimension four \cite{D4}, 
three \cite{D3} and two \cite{D2}. 
Of special interest is the case of D6-branes because they lift to pure 
geometry in eleven dimensions. 
In this note, using eight-dimensional gauged supergravity, which is the 
natural arena to perform twisting for the D6-branes, 
we will briefly review the eleven-dimensional description of 
D6-branes wrapped on 
coassociative 4-cycles \cite{Hernandez}, deformed special Lagrangian 
3-cycles \cite{HS} 
and K\"ahler 4-cycles \cite{GM,HS2}, thus recovering the lifts considered 
in \cite{Gomis}, and study several 
interesting limit cases of the solutions. 
We refer to the original literature for much of the technical details 
and related generalizations (see also 
the contribution by J. Gomis to this proceedings).
  
Maximal gauged supergravity in eight dimensions was constructed 
through Scherk--Schwarz compactification of eleven-dimensional supergravity on 
an $SU(2)$ group manifold \cite{Salam}. The field content in the purely gravitational sector of 
the theory consists of the metric $g_{\mu \nu}$, a dilaton 
$\Phi$, five scalars given by a unimodular $3 \times 3$ matrix $L_{\alpha}^{i}$ in the coset 
$SL(3, \mathbb{R})/SO(3)$ and an $SU(2)$ 
gauge potential $A_{\mu}$, and on the fermion side 
the pseudo--Majorana spinors $\psi_{\gamma}$ and $\chi_i$. We will 
chose as representation 
for  the Clifford algebra
\begin{equation}
\Gamma^{a} = \gamma^{a} \times \II_2 \ ,\qq
 \hat{\Gamma}^{i} = \gamma_9 \times \tau^{i} \ ,
\end{equation}
where as usual $\gamma_9 = i \gamma^0 \gamma^1 \ldots \gamma^7$, 
so that $\gamma_9^2 = \II$, while $\tau^{i}$ 
are Pauli matrices. It will prove useful to introduce 
$\Gamma_9 \equiv \frac {1}{6i} \epsilon_{ijk} \hat{\Gamma}^{ijk} 
= -i \hat{\G}_1 \hat{\G}_2 \hat{\G}_3 = 
\gamma_9 \times \II_2$.
  
\underline{\bf Spin(7) holonomy}: We will first construct a supergravity 
solution corresponding to D6-branes wrapped 
on a coassociative 4-cycle in a seven manifold of $G_2$ holonomy. 
Coassociative 4-cycles 
are supersymmetric cycles preserving $1/16$ supersymmetry. 
Therefore, the solution will be a supergravity 
dual of a three-dimensional gauge theory with $N=1$ supersymmetry 
and when lifted to eleven dimensions will correspond 
to M-theory on an eight manifold with Spin(7) holonomy group 
\cite{Gomis,Hernandez}. The symmetry group of the wrapped branes 
is $SO(1,2) \times SO(4) \times SO(3)_R$. The 
twisting is performed by identifying the structure group of the normal 
bundle, $SO(3)_R$, 
with $SU(2)_L$ in $SO(4) \simeq SU(2)_L \times SU(2)_R$. This leads to a pure gauge theory in three 
dimensions with two supercharges. There are no scalars because the bundle of anti self-dual two-forms is 
trivial; therefore $L^{i}_{\a}=\delta^{i}_{\a}$. The deformation 
on the worldvolume of the D6-brane will be described by the metric ansatz 
\be
ds_8^2 = \a^2 ds_4^2 + e^{2f} dx_{1,2}^2 + d\r^2 \, ,
\label{7}
\ee
with the 4-cycle taken as a 4-sphere with metric de Sitter's metric on $S^4$,
\be
ds_4^2 = \frac {1}{(1 + \xi^2)^2} 
\left( \frac {\xi^2}{4} (\s_1^2 + \s_2^2 + \s_3^2) + d \xi^2 \right),
\label{8}
\ee 
where the left-invariant Maurer--Cartan $SU(2)$ 1-forms satisfy 
$d\s_i= \ha \e_{ijk} \s_j\wedge \s_k$. A useful representation
in terms of the Euler angles $\th$, $\psi$ and $\phi$, is
\be
\s_1\pm i \s_2 = e^{\pm i \psi}(\sin\th d\phi + i d \th)\ ,\qq 
\s_3 = d\psi + \cos\th d\phi \ .
\label{mma}
\ee
We will use for the 8-bein the basis $e^7=d\r$, $e^8=(1+\xi^2)^{-1} d\xi$, etc. From the structure equations, the spin connections on 
$S^4$ can be easily shown to be
\be
\omega_{i \, 8 } = \frac {1 -\xi^2}{1 + \xi^2} \frac {\s_i}{2}\ ,\qq
\om_{ij}=\ha \e_{ijk}\s_k\ .
\label{9}
\ee
The twisting, which amounts to an identification of the spin connection 
with the $R$-symmetry, 
is performed by turning on an $SU(2)$ gauge field obtained from the self-dual 
combinations of the spin connection on $S^4$, 
$A^{1} = - \frac {1}{2} (- \omega_{81} - \omega_{23})$ (+ cyclic). 
The gauge field is then that 
for the charge one $SU(2)$ instanton on $S^4$,
\be
A = \frac {1}{2} \frac {1}{1 + \xi^2} i \, \s_i \tau^{i}.
\label{13}
\ee
Consistency of the Killing spinor equations  
requires the four projections 
\be
\G_{8i} \e =-\ha \e_{ijk} {\hat \G}_{jk} \e \  ,\qq \G^7 \e = - i \G^9 \e\ .
\label{proo}
\ee
These projections lead to Spin(7) invariant Killing spinors since they imply 
the conditions
\be
\Bigl(\G_{\a\b}+{1\ov 6} \Psi_{\a\b\g\d}\G_{\g\d}\Bigr)\e=0\ ,
\label{coon}
\ee
where $\Psi_{\a\b\g\\d}$ is the totally antisymmetric 4-index tensor that is 
invariant under the Spin(7) subgroup of $SO(8)$. To prove that we 
use the standard splitting $\a=(a,8)$, with $a=1,2,\dots , 7$ and denote by 
$\psi_{abc}=\Psi_{abc8}$ the octonionic structure constants. In the basis
$a=(7,i,\hat i=i+3)$, $i=1,2,3$ we have that 
\ba
&&\psi_{ijk}=\e_{ijk}\ ,\qq \psi_{i\hj\hk}=-\e_{ijk}\ ,
\qq \psi_{7i\hj}=\d_{ij}\ ,
\nonumber\\
&& \psi_{7ij\hk} = \e_{ijk}\ ,\qq \psi_{7\hi\hj\hk} =-\e_{ijk}\ ,\qq
\psi_{ij\hat m \hat n}=\d_{im}\d_{jn}-\d_{in}\d_{jm}\ ,
\label{occt}
\ea
and we may easily see that \eqn{proo} imply the conditions
\eqn{coon} for a Spin(7) invariant Killing spinor. 
These projections and the gauge field (\ref{13}) 
lead the Killing spinor equations to
\ba
\frac {df}{d\rho}  =  \frac {1}{3} \frac {d\Phi}{d\r} = 
- \frac {1}{2} \frac {e^{\Phi}}{\a^2} + \frac {1}{4} e^{-\Phi} \ ,\qq
\frac {1}{\a} \frac {d\a}{d\rho}  =  \frac {e^{\Phi}}{\a^2} 
+ \frac {1}{4} e^{-\Phi} \ . 
\label{Killing}
\ea
In terms of a new radial variable defined as $dr = e^{-\Phi/3} d \r$ 
the solution to the differential equations (\ref{Killing}) 
can be lifted to eleven dimensions 
using the elfbein in \cite{Salam}, and leads to \cite{Hernandez}
\be
ds_{11}^2 = ds_{1,2}^2 + \frac {d r^2}{\left( 1 - \frac {l^{10/3}}{r^{10/3}} 
\right)} + \frac {9}{100} r^2 \left( 1 - \frac {l^{10/3}}{r^{10/3}} 
\right) (\Sigma_{i} - A^{i})^2 + \frac {9}{20} r^2 ds_4^2,
\label{20}
\ee
where the $\S_i$'s are the left-invariant Maurer--Cartan
1-forms corresponding to the internal, from an eight-dimensional point of view,
$SU(2)$, for which an 
explicit parametrization similar to \eqn{mma} may be used (with the 
Euler angles denoted by $\th', \psi'$ and $\phi'$).
This is the metric of a Spin(7) holonomy manifold \cite{BS}, 
with the topology of an 
$\mathbb{R}^4$ bundle over $S^4$.
We also note that 
M-theory realizations of the strong coupling description of 
D6-branes wrapped on coassociative cycles involving 
more general Spin(7) manifolds can be identically 
obtained using the non-homogeneous quaternionic spaces of \cite{nonhom}. 
  
\underline{\bf $G_2$ holonomy}: We will now describe how the 
supergravity configuration corresponding to D6-branes wrapped on a deformed 
3-sphere in a Calabi-Yau threefold can be lifted to M-theory 
on a seven manifold of $G_2$ holonomy 
with an $SU(2) \times SU(2)$ isometry group \cite{HS}. 
The wrapped D6-branes on the deformed 3-cycle 
are described by a metric ansatz 
\be
ds_8^2 = \a_1^2 \s_1^2 + \a_2^2 \s_2^2 + \a_3^2 \s_3^2 + e^{2f} 
ds_{1,3}^2 + d\r^2 \, .
\label{s8}
\ee
Deformation of the 3-cycle requires the existence of scalars on the coset manifold, 
$L^i_{\a} = \hbox {diag} ( e^{\l_1}, e^{\l_2}, e^{\l_3})$, with the constraint ${\l_1}+{\l_2}+{\l_3}=0$. 
Within this ansatz, the only consistent way to obtain non-trivial solutions to 
the supersymmetry variations is to impose on the spinor $\e$ the projections
\be
\G_{ij} \e = - \hat{\G}_{ij} \e\ , \: \: \: \: \G^7 \e = - i \G^9 \e\ .
\label{projections}
\ee
Among the possible pairs $\{ij\}=\{12,23,31\}$ only two are independent, so that 
(\ref{projections}) represents three conditions and reduces the number of supersymmetries to $32/2^3=4$, leading 
to $N=1$ in four dimensions.\footnote{In the basis \eqn{occt} the projections
\eqn{projections} imply the condition for a $G_2$ invariant Killing spinor,
\[
\Bigl(\G_{ab}+{1\ov 4} \psi_{abcd}\G_{cd}\Bigr)\e=0\ .
\]
}
The symmetry group of the wrapped branes is now 
$SO(1,3) \times SO(3) \times SO(3)_R$. Supporting covariantly constant spinors on the worldvolume 
of the brane requires again some twisting or mixing of the spin and gauge connections. 
However, in the presence of scalars the twisting is not a direct
identification of the two connections. 
Instead, the generalized twisting in terms of a gauge field 
$A^{\a} = A_i^{\a} \s_i$ is \cite{HS}
\be
A_1^1 = \frac {\a_1}{2} \Big[ - \frac {\omega_1^{23}}{\a_1} \cosh \l_{23} 
+ e^{\l_{21}} \sinh \l_{31} 
\frac {\omega^{31}_2}{\alpha_2} - e^{\l_{31}} \sinh \l_{12} 
\frac {\omega^{12}_3}{\a_3} \Big] \, ,
\label{A1}
\ee 
and $A^2_2$ and $A^3_3$ obtained from cyclicity in the $1,2,3$ indices. We have used the notation 
\be \omega^{jk}_i= \e_{ijk}\frac {\a_j^2 + \a_k^2 - \a_i^2}{2\a_j \a_k}\ ,
\label{sppin}
\ee
for the spin connection along the 3-sphere expanded as 
$\om^{jk}=\om^{jk}_i \s_i$, and $\l_{ij}=\l_i-\l_j$. If we define some new variables, 
$a_i=e^{-\Phi/3} \a_i$, $b_i = e^{2\Phi/3} e^{\l_i}$, $e^{2\Phi}=b_1
b_2 b_3$, $dr= e^{-\Phi/3}  d\rho$,
the Killing spinor equations become
\be
{da_1\ov dr} = - \frac {b_2}{a_3} F^2_{31} - \frac {b_3}{a_2} F^3_{12}\ , 
\qq
{db_1\ov dr} = \frac {b_1^2}{a_2a_3} F^1_{23} -{1 \ov 2 b_2 b_3 } 
(b_1^2 - b_2^2 - b_3^2) \ 
\label{ba}
\ee
and cyclic in the $1,2,3$ indices, where 
the field strength components $F^{i}_{jk} = A_i^{i} + g A_j^{j} A_k^k$ in the $\s^{j} \wedge \s^k$ basis in 
\eqn{ba} are computed using (\ref{A1}) in the new variables,
\be
A_1^1 = - {1\over 2} \frac {d_2^2 + d_3^2 -d_1^2}{2d_2d_3}
\equiv - \frac {1}{2} \Omega^{23}_1\ ,
\label{A11}
\ee
where $d_i \equiv \frac {a_i}{b_i}$ and cyclic in $1,2,3$.
We see that the generalized twist condition \eqn{A1} takes the form 
of the ordinary twist, but for an auxiliary deformed 3-sphere metric 
obtained by replacing the $a_i$'s by the $d_i$'s defined above. 
The lift to eleven dimensions of our eight-dimensional background 
is of the form $ds_{11}^2= ds^2_{1,3}+
ds^2_7$, where 
\be
ds_7^2 = dr^2 + \sum_{i=1}^3 a_i^2 \s_i^2 +
\sum_{i=1}^3 b_i^2(\S_i + c_i \s_i)^2\ ,
\label{s7}
\ee
with $c_i = 2 A_i^{i}$. This metric, when the various functions are subject to the 
conditions \eqn{ba} and \eqn{A11}, describes $G_2$ holonomy manifolds
with an $SU(2)\times SU(2)$ isometry. This can be proved explicitly 
\cite{HS} noting that 
the system of equations (\ref{ba}) can also be derived from self-duality 
of the spin connection for the seven manifold (\ref{s7}), 
since self-duality of the spin 
connection in seven dimensions implies closedness and co-closedness 
of the associative three-form and, therefore, $G_2$ holonomy. 
An extra $SU(2)$ isometry develops when $a_1=a_2=a_3$ and $b_1=b_2=b_3$.
In that case there is no need  
for scalar fields, and the branes wrap a round 3-sphere. 
Then, the system \eqn{ba} simplifies enormously and 
can be solved explicitly \cite{EN}, 
leading naturally to the metric of \cite{BS}. 
  
Let us now consider several interesting limits of the metric (\ref{s7}). 
First we will study the 
case where the radius of the ``spacetime'' 
3-sphere becomes very large so that it can be approximated by 
$\mathbb{R}^3$ and the D6-branes are 
effectively unwrapped. This limit 
can be taken systematically as follows: consider the rescaling $\s_i\to \e 
dx_i$, $b_i\to \e b_i$ and $r\to \e r$ in the limit $\e\to 0$. 
Then, since the functions $c_i=2 A^i_i$ do not scale, the metric 
\eqn{s7} takes the form $ds^2_7=dx_i^2 + ds^2_4$, where we have absorbed a 
factor of $\e^2$ into a redefinition of the overall Planck scale and 
where the four-dimensional non-trivial part of the metric is
\be
ds_4^2 = dr^2 + \sum_{i=1}^3 b_i^2 \S_i^2\ .
\label{EEHH}
\ee 
The system of equations \eqn{ba} reduce to the statement
that the coefficients $a_i$ are constants and therefore they can 
be absorbed into 
a rescaling of the new coordinates $x_i$, 
as we have already done above, and the simpler system
\be
{db_1\ov dr}= {1\ov 2 b_2 b_3}(b_2^2+b_3^2-b_1^2)\ ,
\qq {\rm and\ cyclic\ permutations}\ ,
\label{grt}
\ee
which is nothing but the Lagrange system. 
Four-dimensional metrics \eqn{EEHH} governed by that system 
correspond to a class of hyperk\"ahler 
metrics with $SU(2)$ isometry with famous example, if an extra $U(1)$ 
symmetry develops (for instance when $b_1=b_2$), the 
Eguchi--Hanson metric which is the first non-trivial ALE 
four-manifold.
This is in agreement with the fact that the near horizon limit of 
(unwrapped) D6-branes of 
type IIA when uplifted to M-theory contains, besides the D6-brane 
worldvolume, the Eguchi--Hanson metric.

We will now decouple just one of the coordinates in the cycle.
First we consider the case where $a_1=a_2=a$ and $b_1=b_2$ so that 
an extra $U(1)$ symmetry develops. The resulting simplified system of equations
in \eqn{ba} is still quite complicated and there is no explicit solution to 
it, up to date, leading to regular metrics. Next, let the change of 
variables $\psi'=\varphi + x/\e$, $\psi=x/\e$, followed by setting $a_3=\e$, 
and by taking the limit $\e\to 0$. 
We are then left with a 
five-dimensional field theory with $N=1$ supersymmetry, and the metric 
splits as $ds_{11}^2 = ds_{1,4}^2 + ds_6^2$, where the non-compact 
variable $x$ parametrizes the fifth dimension.
The system \eqn{ba} becomes
\be
\frac {da}{dr} = \frac {1}{2} \frac {b_3}{a} \ ,\qq\frac {db_3}{dr} = 
1 - \frac {1}{2} \frac {b_3^2}{a^2} - \frac {1}{2} \frac {b_3^2}{b_1^2} \ ,
\qq \frac {db_1}{dr} = \frac {1}{2} \frac {b_3}{b_1} \ 
\label{avf}
\ee
and also $c_1=c_2=0$ and $c_3=-1$. 
Hence, we recover the case of D6-branes wrapped on a $\hbox{sLag}_2$ cycle 
(special Lagrangian 2-sphere) studied in \cite{EN}, 
with $ds_6^2$ the metric for the 
resolved conifold \cite{Candelas}. 

\underline{\bf $SU(4)$ holonomy}: D6-branes 
wrapped on a K\"ahler 4-cycle inside a Calabi--Yau threefold 
lift to M-theory on a Calabi--Yau four-fold \cite{Gomis,GM,HS2}. 
The symmetry group of the branes is $SO(1,2) \times SO(4) \times U(1)_R$. 
The R-symmetry is broken to the $U(1)_R$ corresponding to the 
two normal directions to the D6-branes that are inside the 
three-fold. The twisting amounts 
to the identification of this $U(1)_R$ with a $U(1)$ subgroup in one of the $SU(2)$ factors in 
$SO(4)$. The remaining scalar after the twisting, 
the vector and two fermions preserved by the diagonal group of 
$U(1) \times U(1)_R$ determine the field content of $N=2$ 
three-dimensional Yang--Mills. Following \cite{HS2} we take
the 4-cycle to be a product of 
two 2-spheres of different radii, $S^2 \times {\bar S}^2$. The
worldvolume of the D6-branes will be described by a metric 
\be
ds_8^2 = e^{2f} ds_{1,2}^2  + d\r^2 + 
\a^2  d\Om_2^2 + \b^2  d{\bar\Om}_2^2\ .
\label{metric}
\ee
Consistency of the Killing spinor equations 
requires turning on one of the scalars in the coset, 
$L^{i}_{\a} = \hbox {diag } ( e^{\l}, e^{\l}, e^{-2 \l})$, and one of the components of the gauge field,
\be
A^3 =  - \frac {1}{2}(\s_3+{\bar \s}_3)\ , 
\label{wkjh}
\ee
thus realizing the breaking of the $SU(2)_R$ 
R-symmetry to $U(1)_R$, and the projections 
\be
\G_1 \G_2 \e =  \bar{\G}_1 \bar{\G}_2 \e = - \hat{\G}_1 \hat{\G}_2 \e, \: \: \: \: 
\G_7 \e =  - i \G_9 \e \ .
\label{proj}
\ee
These leave in total four independent components for the spinor. If we redefine variables as 
$dr = e^{-\Phi/3} d \r$, $a = e^{- 2 \l + 2 \Phi/3}$, $a_1 = \a \, e^{-\Phi/3}$, 
$a_2 = \beta e^{- \Phi/3}$, $a_3   =  e^{\l + 2 \Phi/3}$, 
the Killing spinor equations become
\be
\frac {d a}{dr} = 1- \frac {a^2}{2 } \left({1\ov a_1^2} 
+ {1\ov a_2^2} + {1\ov a_3^2}\right) \ ,\qq
a_1 \frac {da_1}{dr} = \frac {a}{2}\ ,\qq  {\rm and\ cyclic\ in\ 1,2,3}\ .
\label{ABCD}
\ee
The general solution to the system \eqn{ABCD} is 
\be
a_1^2 =  R^2 + l_1^2, \qq a_2^2 = R^2 + l_2^2 \ , \qq a_3^2 =  R^2, \qq a^2 = R^2 U^2(R) \ ,
\label{zero}
\ee
where
\be
U^2(R) = \frac {3 R^4 + 4 (l_1^2 + l_2^2) R^2 + 6 l_1^2 l_2^2+ 3 C/R^4}
{6 (R^2 + l_1^2 ) ( R^2 + l_2^2 )} 
\label{urr}
\ee
and the two variables $r$ and $R$ relate via the differentials $dr=2 dR /U(R)$. 
Here we have denoted three of the 
constants of integration by $l_1,l_2 $ and $C$
and we have absorbed the fourth one by an appropriate shift in the variable 
$R$. When lifted to eleven dimensions the solution factorizes 
into the three-dimensional flat space-time 
and a Calabi--Yau four-fold
\be
ds_{11}^2 = ds_{1,2}^2 +
\frac {4 dR^2}{U^2(R)}  + a_1^2 d\Om^2_2 
+  a_2^2 d{\bar \Om}^2_2 + a_3^2 d{\hat \Om}^2_2  
+ a^2 \left( \hat{\s}_3 - \s_3 - \bar{\s}_3 \right)^2 \ .
\label{lift}
\ee
Topologically the Calabi--Yau four-fold is a 
$\mathbb{C}^2$ bundle over $S^2 \times S^2$ and 
was also constructed with a different method in \cite{Cvetic2} (for $C=0$).
This result includes those obtained in \cite{GM,Gursoy}, 
where both spheres in the four-cycle were taken to have the same radius, 
so that $l_1=l_2$. Having unequal radii makes possible to consider the 
limit where the radius of one of the three spheres tends to infinity.
Indeed, if we take the limit $l_2\to \infty$, we have that 
$a_2^2 d\bar \Om^2_2 =dx_1^2+dx_2^2$, i.e. $\bar S^2\to \mathbb{R}^2$,
and the metric \eqn{lift} becomes
$ds^2_{1,4} + ds_6^2$, where $ds_6^2$ is the metric of the resolved conifold
\cite{Candelas} in its standard form, up to a factor of 6 and after letting $l_1=\sqrt{6} a$ and $C=0$.
Note also that, in this 
limit, the systems \eqn{ABCD} and \eqn{avf}, after an 
appropriate renaming of the variables, coincide, as it should be.


\bs
 
\centerline {\bf Acknowledgments}

The authors acknowledge the financial support provided by the European
Community's Human Potential Programme under contract HPRN-CT-2000-00131
Quantum Spacetime.
R.H. also acknowledges the Swiss Office for Education and Science 
and the Swiss National Science Foundation, 
and the organizers of the Leuven workshop for a 
stimulating atmosphere. K.S. acknowledges the financial support 
provided through the European
Community's Human Potential Programme under contract 
HPRN-CT-2000-00122 Superstring Theory, by the Greek State
Scholarships Foundation under the contract IKYDA-2001/22, 
as well as NATO support by a Collaborative Linkage Grant 
under the contract PST.CLG.978785. 

\bs

\underline{\em Note added:} After completion of this note, we learned from reference \cite{twistagain}, where 
the construction in \cite{HS} was extended to include all metrics of $G_2$ holonomy with 
$S^3 \times S^3$ principal orbits through an extension of the twisting procedure.


\end{document}